# Multimessenger Exoplanets: Combining Infrared Excess and Astrometry with Gravitational-wave Detections of Double White Dwarf Exoplanets

Casey McGrath[1,2,3], Mary Anne Limbach[4], Scott C. Noble[3], and Vincent Kofman[5]
[1] Center for Space Sciences and Technology, University of Maryland, Baltimore County, Baltimore, MD 21250, USA; casey.d.mcgrath@nasa.gov
[2] Center for Research and Exploration in Space Science and Technology II, NASA/GSFC, Greenbelt, MD 20771, USA
[3] Gravitational Astrophysics Lab, NASA Goddard Space Flight Center, Greenbelt, MD 20771, USA
[4] Department of Astronomy, University of Michigan, Ann Arbor, MI 48109, USA
[5] Centre for Planetary Habitability, Department of Geoscience, University of Oslo, Oslo, Norway

## Abstract

Multimessenger astrophysics provides a powerful framework for characterizing astrophysical systems through multiple, independent signals rather than relying solely on electromagnetic (EM) observations. While most prior multimessenger studies have focused on the properties of ultracompact objects, recent work has shown that gravitational waves (GWs) detectable with the upcoming Laser Interferometer Space Antenna (LISA) observatory can also be used to detect exoplanets orbiting stellar remnants. In particular, this includes circumbinary planets around tight double white dwarf (DWD) systems. This population is unique: no such planets are currently known around DWDs, and only a handful of circumbinary planets around main-sequence stars and planets orbiting white dwarfs (WDs) have been identified. Independent EM confirmation will be essential, especially for the first GW-detected exoplanets, to validate this method. While EM techniques are unlikely to efficiently survey large populations of DWDs, GW detection followed by targeted EM confirmation provides one pathway for identifying and characterizing these systems. Using a representative case, the known DWD ZTF J2243+5242 scaled to a fiducial distance of 150 pc, we compute the detectability of circumbinary exoplanets via LISA GW observations. We then demonstrate two complementary EM techniques—JWST infrared (IR) excess and Gaia or Habitable Worlds Observatory astrometry—that could independently confirm a GW-detected planet in this fiducial system. We present the first exploration of IR excess around DWDs as a viable exoplanet detection method, extending its established use for solitary WDs. Both EM techniques overlap with regions of parameter space where LISA is sensitive to DWD exoplanets, enabling robust confirmation and characterization.

## 1. Introduction

In the past decade, we have seen the birth of the field of observational gravitational-wave (GW) astrophysics, which has discovered hundreds of transient GW events to date, mostly originating from the mergers of binary black holes (A. G. Abac et al. 2025). GW observatories were originally envisioned for studying the exotic remnants of stars—black holes, neutron stars, and white dwarfs (WDs). More recent studies (T. Robson et al. 2018; N. Tamanini & C. Danielski 2019; W.-L. Guo et al. 2025) have shown it may also be possible to detect and study stellar objects or even exoplanets orbiting these same compact objects. N. Seto (2008) first described the idea of the GW radial velocity (RV) method for detecting exoplanets, with the focus on how the upcoming Laser Interferometer Space Antenna (LISA) mission could potentially discover circumbinary exoplanets orbiting double white dwarfs (DWDs).[6] As the development of LISA has progressed more attention has gone toward developing models and detection techniques for finding the signals of these exoplanets in the future LISA datasets (C. Danielski et al. 2019; M. L. Katz et al. 2022; C. McGrath 2026), as well as population studies to try and understand their possible formation channels (G. Columba et al. 2023; S. Ledda et al. 2023; A. Nigioni et al. 2026).

Despite the thousands of exoplanets discovered around main-sequence stars, less than 10 have been detected around WDs (e.g., see D. Veras 2021). Yet this population is key to understanding the ultimate fate of most planetary systems, as the majority of stars, including the Sun, end their lives as WDs. Depending on the stellar mass, planets further out than ~2 au are expected to survive post-main-sequence evolution largely intact (J. Nordhaus & D. S. Spiegel 2013). And indeed, a handful of WD planetary systems and candidates have been reported beyond this putative survival boundary (S. E. Thorsett et al. 1993; S. Sigurdsson et al. 2003; K. L. Luhman et al. 2011; J. W. Blackman et al. 2021). Discoveries near or interior to this boundary are especially informative for post-main-sequence dynamical evolution and migration. A few such systems have now been identified (B. T. Gänsicke et al. 2019; A. Vanderburg et al. 2020), suggesting that some planets can settle into tight WD orbits.

The scarcity of WD exoplanet detections does not necessarily imply a low occurrence rate. Rather, it reflects the challenges of applying the dominant discovery methods for main-sequence hosts—RV Doppler monitoring and transits—to compact stellar remnants. These techniques are not optimally suited to WDs, due to their intrinsic faintness, small radii, and lack of prominent spectral features (M. A. Limbach et al. 2022; L. K. Rogers et al. 2024).

[6] This work focuses on LISA as the primary space-based GW observatory, but we also note that similar missions such as Taiji or TianQin could also enable these discoveries (Z. Luo et al. 2020; E.-K. Li et al. 2025). For example, Y. Kang et al. (2021) compare the detection prospects of DWDs between Taiji and LISA.

With JWST, this picture is changing. Early surveys targeting nearby WDs have nearly doubled the number of candidates identified via direct imaging and infrared (IR) excess (M. A. Limbach et al. 2024; S. E. Mullally et al. 2024; J. H. Debes et al. 2025; S. Poulsen et al. 2026), although some false positives are expected (F. Mullally et al. 2026). However, even if these candidates are confirmed (M. A. Limbach et al. 2025), the discovery channels provide direct atmospheric characterization but limited or no independent constraints on orbits, and they poorly constrain physical parameters such as mass. Additional techniques capable of independent confirmation and robust orbital/physical parameters are needed to build a complete demographic picture.

GW observations offer complementary sensitivity to circumbinary planets orbiting DWDs, a regime with no confirmed planets to date. However, there is some compelling evidence that such planets may exist. Not only are there the aforementioned WD exoplanets, but there are circumbinary planets known around main-sequence binaries (e.g., L. R. Doyle et al. 2011; W. F. Welsh et al. 2015; V. B. Kostov et al. 2020; P. Thebault & D. Bonanni 2025). Beyond providing independent confirmation and precise orbital characterization, detecting planets around DWDs would illuminate the prevalence and survival of circumbinary planets and how these systems evolve after stellar death, paralleling studies of planets around solitary WDs (D. Veras 2021).

In this study, we explore the *multimessenger* detectability of exoplanets around DWDs by coupling electromagnetic (EM) and GW detection methods. Our aim is to map the overlap in parameter space where EM and GW techniques complement each other, and to assess how joint detections deliver additional constraints on orbital and physical exoplanet properties. We focus on astrometry and IR excess as the most promising EM techniques for confirming LISA-detected DWD exoplanets, as other methods show little overlap with the GW exoplanet detection parameter space (N. Tamanini & C. Danielski 2019). This work also continues to explore fundamental ideas and concepts behind the GW RV method for exoplanet detection, as presented in C. McGrath (2026). As such, we will point to that study throughout this work. Note that only planets orbiting *both* WDs ("p-type" planets) are considered because the WDs need to be extremely close to each other in order to generate GWs (as opposed to an "s-type" configuration where the planet is orbiting only one of the WDs). And in this configuration, the planet could gravitationally tug on the combined center of mass of the system, resulting in the Doppler modulation of the observed GW signal.

This paper explores the GW detection of DWD exoplanets and viable methods for their EM confirmation through a representative case study. Specifically, we adopt the parameters of a known DWD system, place it at a fiducial distance of 150 pc, and examine the detectability space for GW observations and the two EM techniques considered. In Section 2, we evaluate GW RV sensitivity to exoplanets in this hypothetical system, and then compute the corresponding JWST IR excess sensitivity. Section 3 presents the joint detection capabilities of these methods, and includes additional estimates for astrometric detection with the Habitable Worlds Observatory (HWO) and Gaia missions. Section 4 discusses important caveats, primarily for the GW detection method with LISA, and Section 5 summarizes our findings. The Appendix presents the details for our estimates of the astrometric sensitivity.

## 2. Methods

In this section, we introduce the system we use for our test case (ZTFJ2243+5242) and the methods used to determine GW RV and IR excess detectability of DWD exoplanets.

### 2.1. The ZTFJ2243+5242 Test Case

To demonstrate the detectability of exoplanets using LISA GW observations, we begin by motivating a hypothetical DWD system for our case study. As a GW observatory, LISA will operate in the $10^{-4}$–$10^{-1}$ Hz part of the GW spectrum; the "mHz" GW regime (M. Colpi et al. 2024). LISA's greatest sensitivity will fall around 1–10 mHz, which corresponds to binaries with orbital periods of ∼3–30 minutes (note that assuming circular orbits for the WDs, the GW frequency is twice the DWD orbital frequency, $f_{\mathrm{GW},0} = 2/P_{\mathrm{DWD}}$). This is where most DWD systems are expected to be individually distinguishable in the GW signal. Below approximately 1 mHz, the density of DWD signals in frequency space becomes so high that individual sources blend into an unresolved astrophysical foreground (i.e., confusion noise). It is referred to as a "foreground" noise because the Galactic DWDs sources that LISA will find will be closer than the other types of more distant GW sources beyond the Milky Way that LISA will also discover.

EM observations have identified only a small fraction of the total DWD population. A compilation of these systems, along with their electromagnetically measured parameters and predicted LISA signals, is presented in E. Finch et al. (2023). LISA is expected to discover a much larger population of DWDs, as its sensitivity is not limited to eclipsing systems.

To assess the GW detectability of circumbinary planets, we adopt a representative DWD system based on ZTFJ2243+5242, relocated from its measured distance $D$ of 2425 pc to 150 pc, consistent with expectations that LISA will detect orders of magnitude more DWDs, including a nearby population. Additional fiducial values for the binary inclination angle $\iota$, polarization angle $\psi$, and initial phase $\phi_0$ are indicated. The adopted parameters for the WD masses $M_A$ and $M_B$, orbital period $P_{\mathrm{DWD}}$, and sky angles $\lambda$ and $\beta$ are summarized in Table 1, based on measurements taken by K. B. Burdge et al. (2020).[7]

We adopt a "ZTFJ2243-like" DWD system as a representative case, as its frequency lies outside the regime of astrophysical foreground noise, and it is predicted to produce a relatively loud signal (E. Finch et al. 2023). Placing this system at a closer distance further increases its signal strength.

### 2.2. GW Exoplanet Discovery Space

To calculate the signal strength of a GW source, we compute the signal-to-noise ratio (SNR) relative to the

[7] Note that T. Kupfer et al. (2024) report the distance of ZTFJ2243+5242 as 1756 pc. This distance and that measured by K. B. Burdge et al. (2020) are consistent within their uncertainty bars, but the reason for their difference is due to different methods of measuring the distances (direct versus indirect).

**Table 1**
Summary of ZTFJ2243-analog System Parameters

| Parameter | Value | Unit | Source |
|---|---|---|---|
| ZTFJ2243 Measured Parameters | | | |
| $M_A$ | 0.323 | $M_\odot$ | B20 |
| $M_B$ | 0.335 | $M_\odot$ | B20 |
| $P_{\rm DWD}$ | 527.93 | s | B20 |
| $\lambda$ | 13.24 | deg | B20[a] |
| $\beta$ | 53.96 | deg | B20[a] |
| Fiducial System Properties | | | |
| $D$ | 150 | pc | This work |
| $\iota$ | 90[b] | deg | This work |
| $\psi$ | 0 | deg | This work |
| $\phi_0$ | 0 | deg | This work |
| Derived System Properties | | | |
| $A_{\rm binary}$ | $1.2 \times 10^{-21}$ | ⋯ | This work[c] |
| $f_{\rm GW,0}$ | 3.79 | mHz | This work |
| $\dot{f}_{\rm GW,0}$ | 3.01 | nHz yr$^{-1}$ | This work[d] |
| SNR (4 yr[e]) | 2310 | ⋯ | This work |
| SNR (7 yr) | 3055 | ⋯ | This work |
| SNR (10 yr[e]) | 3652 | ⋯ | This work |

**Notes.** B20: K. B. Burdge et al. (2020); M26: C. McGrath (2026).
[a] Converted to ecliptic coordinates.
[b] Edge-on to line-of-sight, maximal Doppler effect.
[c] Equation (10), M26.
[d] Equation (A1), M26. Assumes evolution only from GW radiation.
[e] Nominal (extended) LISA mission: 4 (10) yr (M. Colpi et al. 2024).

expected LISA noise floor:

$$\text{SNR}^2 = 4\ \text{Re}\int_0^\infty \frac{|\tilde{M}(f)|^2}{S_N^{(1)}(f)}df, \tag{1}$$

where $S_N^{(1)}$ is a model of the one-sided power spectral density (PSD) of LISA's noise, and $\tilde{M}$ is the signal model in the frequency domain. For simplicity, we model LISA and its noise PSD following the approach presented in C. McGrath (2026). We use the same GW waveforms (Equations (6), (11), and (12) of that work) to generate the "second-generation Michelson" time-delay interferometry (TDI) observable, denoted $X_{2.0}$, using codes from the LISA Simulation Suite (see *Software* in the Acknowledgments). This $X_{2.0}$ observable represents the fractional laser frequency fluctuation, $\Delta f_{\rm laser}/f_{\rm laser}$, and is therefore dimensionless.

The total SNRs reported in Table 1 are calculated using only the $X_{2.0}$ LISA data channel. In reality, LISA will produce multiple independent data channels that contribute to the SNR calculation. For this study, we use a single data channel, which provides a representative estimate of the signal magnitude. A major advantage of GW observatories is their reach—GW amplitudes scale inversely with source distance, $D$, rather than with $D^2$. Therefore, LISA is expected to detect of order $\mathcal{O}(10^4)$ individually resolvable DWD systems throughout the Milky Way (K. Breivik et al. 2020).

Several important aspects of the *total* SNR shown in Table 1 are worth noting:

1. The total SNR of the system is the same for purely monochromatic signals and for signals that include frequency evolution ("chirping") only in the phase. For these LISA DWD systems, the secular change in GW amplitude due to chirping is negligible (see the discussion of Equation (12) in C. McGrath 2026). Therefore, the total radiated power remains unchanged; only its distribution over time in the observed GW spectrum is altered.
2. The total SNR of the system also remains the same whether or not an exoplanet exists in a circumbinary configuration. The Doppler effect induced by the exoplanet does not alter the total radiated power of the GW system; it only changes how that power is distributed in the observed GW spectrum.[8]
3. For a monochromatic signal with amplitude $A$, frequency $f_0$, and specified noise PSD, observed over a time $T_{\rm obs}$, the total SNR is

$$\text{SNR} \approx A\sqrt{\frac{T_{\rm obs}}{S_N^{(1)}(f_0)}}. \tag{2}$$

The observed LISA GW amplitude does depend on the observatory's antenna pattern, but in general is proportional to the GW strain amplitude $A_{\rm binary}$ (see Equation (12) of C. McGrath 2026). Therefore,

$$\text{SNR} \propto \frac{\mathcal{M}^{5/3}T_{\rm obs}^{1/2}}{D}\frac{f_{\rm GW,0}^{2/3}}{S_N^{(1)}(f_{\rm GW,0})^{1/2}}, \tag{3}$$

where $\mathcal{M} \equiv (M_A M_B)^{3/5}/M_{\rm DWD}^{1/5}$ is the "chirp mass" of the DWD system and $M_{\rm DWD} = M_A + M_B$ is the total system mass. This scaling with observation time $T_{\rm obs}$, is reflected in the total SNR values shown in Table 1.

To explore the exoplanet discovery space of a DWD circumbinary, we compute a "residual SNR" between models with and without an exoplanet:

$$\text{SNR}_{\rm res}^2 = 4\ \text{Re}\int_0^\infty \frac{|\tilde{M}_{\rm exo}(f) - \tilde{M}_{\rm no\ exo}(f)|^2}{S_N^{(1)}(f)}df, \tag{4}$$

where the subscript "exo" denotes the model with the exoplanet, and "no exo" denotes the model without the exoplanet, respectively. In a typical Bayesian analysis, we define and evaluate a likelihood function, $\mathcal{L}$, which relates to the residual SNR between the observed data and a model via $\ln\mathcal{L} \propto -\text{SNR}_{\rm res}^2/2$. Therefore, Equation (4) provides a framework for comparing the two models and quantifying the impact of an exoplanet on the observed signal, without requiring a full Bayesian analysis. Here, we take the residual SNR as a proxy for the companion SNR.

In doing so, we assume that systematics are fully accounted for and that the DWD parameters are known perfectly a priori, allowing fits with and without the companion (as required to compute the Bayesian evidence and determine whether the DWD-only model or the DWD+companion model is favored). However, this optimistic assumption implies that the resulting companion detection limits represent lower bounds that are unlikely to be achievable with LISA in practice.

[8] This was a key finding in Section 3.4 of C. McGrath (2026) that was derived analytically and using a method that was agnostic of the GW observatory. Here we confirm this finding using a full model of the LISA observatory for the case of study presented.

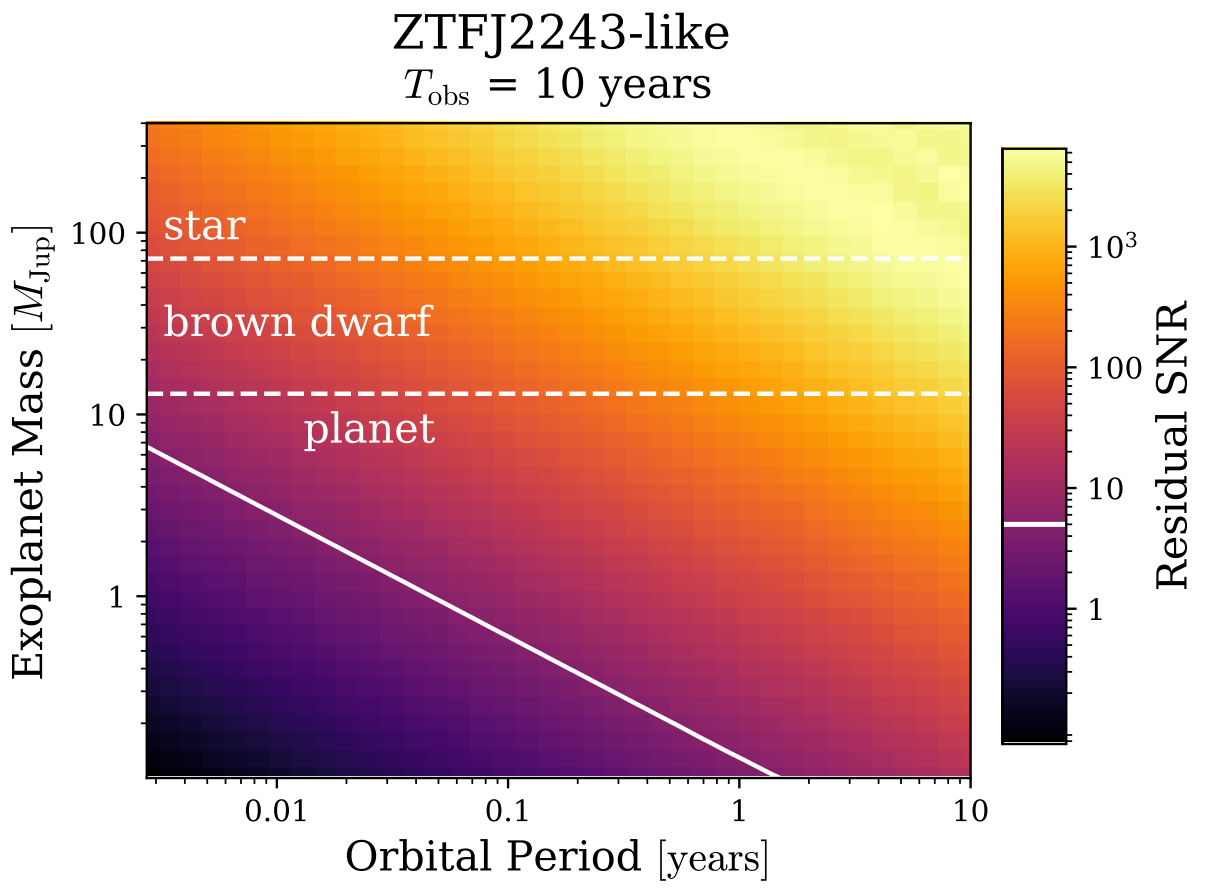

**Figure 1.** LISA exoplanet discovery space for the ZTFJ2243-like system. The residual SNR between models with and without an exoplanet is shown for the ZTFJ2243-like system with a 10 yr LISA observation across a range of exoplanet masses and periods. The exoplanet's inclination is set to $i = \pi/2$ (maximal Doppler effect). The total SNR of the system itself is 3652 for the full 10 yr observation. Dashed lines indicate the boundaries between planets ($M_p \lesssim 13\, M_{\rm Jup}$), brown dwarfs ($13\, M_{\rm Jup} \lesssim M_p \lesssim 72\, M_{\rm Jup}$), and stars ($M_p \gtrsim 72\, M_{\rm Jup}$). The SNR = 5 detection threshold is shown as a solid line.

For the ZTFJ2243-like system, Figure 1 shows the residual SNR for exoplanetary companions across the discovery space. At 10 yr observation, the total SNR is 3652 (see Table 1). Under the (optimistic) assumption that the residual SNR approximates the companion SNR, a single Jupiter-mass exoplanet is detectable over orbital periods of $\sim$0.05–10 yr with an SNR > 5.

We find that the simulated data in Figure 1 are consistent with the power-law relation

$$\left(\frac{P}{1\ \rm yr}\right) \approx 0.07\left(\frac{\rm SNR_{res}}{5}\right)^{3/2}\left(\frac{M_p}{M_{\rm Jup}}\right)^{-3/2} \times \left(\frac{M_{\rm DWD}+M_p}{M_\odot}\right), \quad (5)$$

where $P$ and $M_p$ are the exoplanet period and mass. The pre-factor of 0.07 is specific to our ZTFJ2243-like system and depends on quantities such as the WD masses, the system's distance, and its inclination.

It is useful to compare with C. McGrath (2026), who derived an analytic expression for the fractional power removed from the GW carrier frequency (their Equation (17)) by the exoplanet. The trend shown in their Figure 3 is qualitatively similar to the residual SNR in Figure 1. While the two quantities differ, Equation (17) captures the power removed from the carrier, whereas Equation (4) reflects the redistributed power across the full spectrum. For small values of the "phase modulation index" parameter $\epsilon$ (Equation (7) of C. McGrath 2026), $1 - J_0^2(\epsilon) \approx \epsilon^2/2$. So if the redistributed power across the spectrum comes from the power removed from the GW carrier frequency, then this implies that the residual SNR scales with the phase modulation index as $\mathrm{SNR}_{\rm res} \sim \sqrt{1 - J_0^2(\epsilon)} \sim \epsilon$. This regime is expected for the LISA DWD exoplanet systems considered here.

Additionally, we find that the residual SNR follows the same dependence on distance and observation time as given in Equation (3). We expect the additional scalings with chirp mass, GW frequency, and noise PSD in the residual SNR as well, since these are all common factors in Equations (1) and (4). Combining these scalings yields the generalized relation $\mathrm{SNR}_{\rm res} \propto \epsilon A_{\rm binary}\sqrt{T_{\rm obs}/S_N^{(1)}(f_{\rm GW,0})}$, which can be expressed explicitly as

$$\mathrm{SNR_{res}} \approx 29\left(\frac{P}{1\ \rm yr}\right)^{2/3}\left(\frac{M_p}{M_{\rm Jup}}\right)\left(\frac{M_{\rm DWD}+M_p}{M_\odot}\right)^{-2/3} \times \left(\frac{\mathcal{M}}{0.5\,M_\odot}\right)^{5/3}\left(\frac{150\ \rm pc}{D}\right)\left(\frac{T_{\rm obs}}{4\ \rm yr}\right)^{1/2} \times \left(\frac{f_{\rm GW,0}}{1\ \rm mHz}\right)^{5/3}\left(\frac{S_N^{(1)}(f_{\rm GW,0})}{S_N^{(1)}(1\ \rm mHz)}\right)^{-1/2}\sin(i), \quad (6)$$

where again the pre-factor of 29 is specific to this ZTFJ2243-like system. Fixing the exoplanet inclination, the DWD chirp mass, distance, and GW frequency, and the observation time to the values used in Figure 1 recovers Equation (5). This expression gives us a useful tool for predicting how variations in the exoplanet period, mass, and inclination, as well as the DWD's frequency, distance, and chirp mass, and the observation time affect the LISA exoplanet discovery space shown in Figure 1.

### 2.3. IR Excess with JWST

LISA will be highly effective at surveying for and identifying DWD exoplanet candidates. However, EM follow-up, particularly for the first detections, given that GW-based exoplanet discovery is untested, will be necessary for confirmation. While previous work has identified EM techniques capable of probing similar parameter space (N. Tamanini & C. Danielski 2019), these have not extended to IR excess detection, whose feasibility we explore here.

#### 2.3.1. IR Excess Detectability of Exoplanets around DWDs

To assess the detectability of exoplanets in DWD systems using the IR excess technique with JWST/MIRI broadband imaging, we adopt a forward modeling approach that leverages the unique mid-IR characteristics of planetary thermal emission. Our method builds upon the IR excess technique introduced in M. A. Limbach et al. (2022), where the presence of a cool giant planet is inferred via photometric IR excess relative to a stellar spectral energy distribution (SED) model. This technique has been successfully used to detect exoplanetary IR excess in several recent studies (M. A. Limbach et al. 2024, 2025; J. H. Debes et al. 2025). While previous work considered single WD hosts, we extend this framework to systems containing two spatially unresolved WDs, representing, to our knowledge, the first analysis of IR excess detectability for such DWD configurations.

We begin by modeling the composite spectrum of the DWD system, assuming both WDs emit as ideal blackbodies. This is a reasonable approximation in the mid-IR, especially beyond $\sim$10 $\mu$m, where the WD emission is firmly in the Rayleigh–Jeans tail, leading to a predictable and smooth continuum that can be precisely characterized (M. Kilic et al. 2008).

To this combined DWD blackbody spectrum, we add model spectra of giant exoplanets. We utilize the publicly available Sonora-Bobcat models (M. Marley et al. 2021), which are computed using a 1D radiative-convective equilibrium atmosphere code assuming "rainout" chemistry (M. S. Marley et al.

1999; D. Saumon & M. S. Marley 2008; C. V. Morley et al. 2012, 2014). These spectra, unlike the WDs, are not in the Rayleigh–Jeans tail at JWST/MIRI wavelengths and exhibit broad molecular absorption features, primarily from $CH_4$, $H_2O$, CO, and $NH_3$, that distinguish them from the smooth continuum of the WDs. As such, even modest planetary fluxes can result in significant fractional excess at 10–20 $\mu$m (K. D. Gordon et al. 2022).

We calculate the flux excess introduced by each planet model in the context of the DWD+planet composite system. We do this by computing the system's total SED and then evaluating the resulting flux in each of the MIRI imaging bands. Fluxes in each band are compared to expectations from the DWD-only model to quantify the fractional IR excess introduced by the planet.

In contrast to spatially resolved imaging, which only requires the exoplanet flux to exceed the detection threshold, unresolved detection via IR excess depends on relative flux contributions and system calibration. While spatially resolved detection of exoplanets is a viable technique for very nearby systems, and a reasonable consideration when searching for planets around solitary WDs, the same is not generally true for DWD systems. DWDs with extremely short orbital periods, such as those detectable with LISA, are expected to be intrinsically rare relative to the population of solitary WDs or WDs with wider orbit stellar companions, and thus the majority of detectable systems are anticipated to lie at distances of several hundred parsecs, rather than within a few tens of parsecs (e.g., see K. B. Burdge et al. 2020, which identified 15 such DWDs all of which were beyond 100 pc). At such distances, the angular separation between a giant planet and its DWD host is typically far below the resolution limit of JWST/MIRI (or even forthcoming Extremely Large Telescopes). Consequently, unresolved photometric detection via mid-IR excess is likely the only practical method for identifying giant planets in these systems, provided the planet has sufficient intrinsic luminosity to contribute measurably to the system's total flux.

Note that because planets and WDs have similar brightness in the mid-IR, it is possible to *spatially resolve* planets within the diffraction limit (at only 1–2 pixels of separation) with MIRI (M. A. Limbach et al. 2024; C. Adelman et al. 2025). For DWDs at 100 pc (and with MIRI, which has a plate scale of $0\farcs11$ pix$^{-1}$), spatially resolving exoplanets from their DWD hosts would require projected orbital separations of at least 11 au, with larger planet-DWD orbital separations for systems at further distances. Planets at projected orbital separations $<$11 au can still be detected through spatially *unresolved* IR excess, as considered in this manuscript. Because direct imaging sensitivity is comparable (or better) than the IR excess sensitivities, MIRI may detect planets smaller than those reported here when they are sufficiently separated from their DWD hosts to be spatially resolved.

IR excess detectability is assessed based on two criteria: photometric sensitivity and calibration precision. We can estimate the detectability by looking at the performance of JWST on both these issues. MIRI has a nominal 2% absolute flux calibration, based on JWST calibration standards that include DA WDs and A-type stars (K. D. Gordon et al. 2022, 2025). This 2% floor sets the minimum differential photometric precision achievable when comparing fluxes across bands. To confidently detect a photometric excess, we require at least an SNR = 3 signal, corresponding to a minimum of 6% flux excess. Sensitivity thresholds are determined using MIRI's published performance metrics, which account for photon noise, thermal background limitations, etc. Because the detection limits for IR excess with MIRI have been established observationally rather than theoretically (as is the case for the LISA GW detection limit analysis), we adopt a less conservative threshold of SNR = 3 instead of SNR = 5, since SNR = 3 detections have already been demonstrated.

This analysis framework enables us to evaluate which exoplanet models would result in detectable IR excess, given the DWD system distance and assumed observational integration time. The addition of a second WD in our DWD systems introduces additional flux in the denominator of the IR excess ratio, modestly decreasing the planet's fractional contribution.

### 2.3.2. IR Excess Detection Limits

Using the method described, we computed MIRI's sensitivity to exoplanets across a range of masses for the ZTFJ2243-like system, to find the minimum mass detectable (at SNR = 3) with IR excess. We note that here it is assumed that the emission from the planet is due to heat left over from the planet's formation, not from stellar insolation, and therefore a planet that emits enough flux to enable detection will be detectable at any orbital separation, out to the edge of the field of view (generally thousands of astronomical units).

Figure 2 shows the sensitivity of this method to a 7 $M_{\rm Jup}$ exoplanet in the ZTFJ2243-like system. The left panel illustrates the sensitivity of the various MIRI bands (black bars); any source whose SED lies above a given bar is detectable if the source is in isolation. This demonstrates that, in isolation, MIRI imaging would be capable of detecting this 7 $M_{\rm Jup}$ exoplanet in all MIRI filters with wavelengths shorter than 18 $\mu$m. However, the planet orbits the DWD and is therefore not an isolated point source. In addition to sufficient sensitivity to detect the source, there must also be enough excess flux from the planet above that of the DWD, as discussed in the previous section, to induce at least a 6% excess. From the right panel, we see that such excesses are only achieved for wavelengths longer than 14 $\mu$m. Hence, in this case, the planet is detectable with MIRI, but only using the F1280W (12.8 $\mu$m), F1500W (15 $\mu$m), and F1800W (18 $\mu$m) filters, where both conditions are met and there is sufficient sensitivity and sufficient exoplanet-induced excess above the DWD to enable detection.

Repeating this analysis over a range of exoplanet models spanning different masses, we find that the minimum detectable exoplanet mass for this system is 5 $M_{\rm Jup}$ at SNR = 3, or 7 $M_{\rm Jup}$ at SNR = 5. We note that this calculation relies on a specific set of exoplanet evolutionary models, which in turn requires us to assume an age for the exoplanet. Here we have assumed that the planet has the same age as the WD (0.2 Gyr), although this is strictly a lower limit on the planet's age. For single WDs, it is often possible to obtain a reasonable constraint on the total system age from WD evolutionary models. In the case of DWDs, however, past mass transfer likely alters the evolutionary history of the stars, making the total system age difficult to constrain. If the planet is significantly older than the WD, the detectable mass limits will be correspondingly higher than those we compute here.

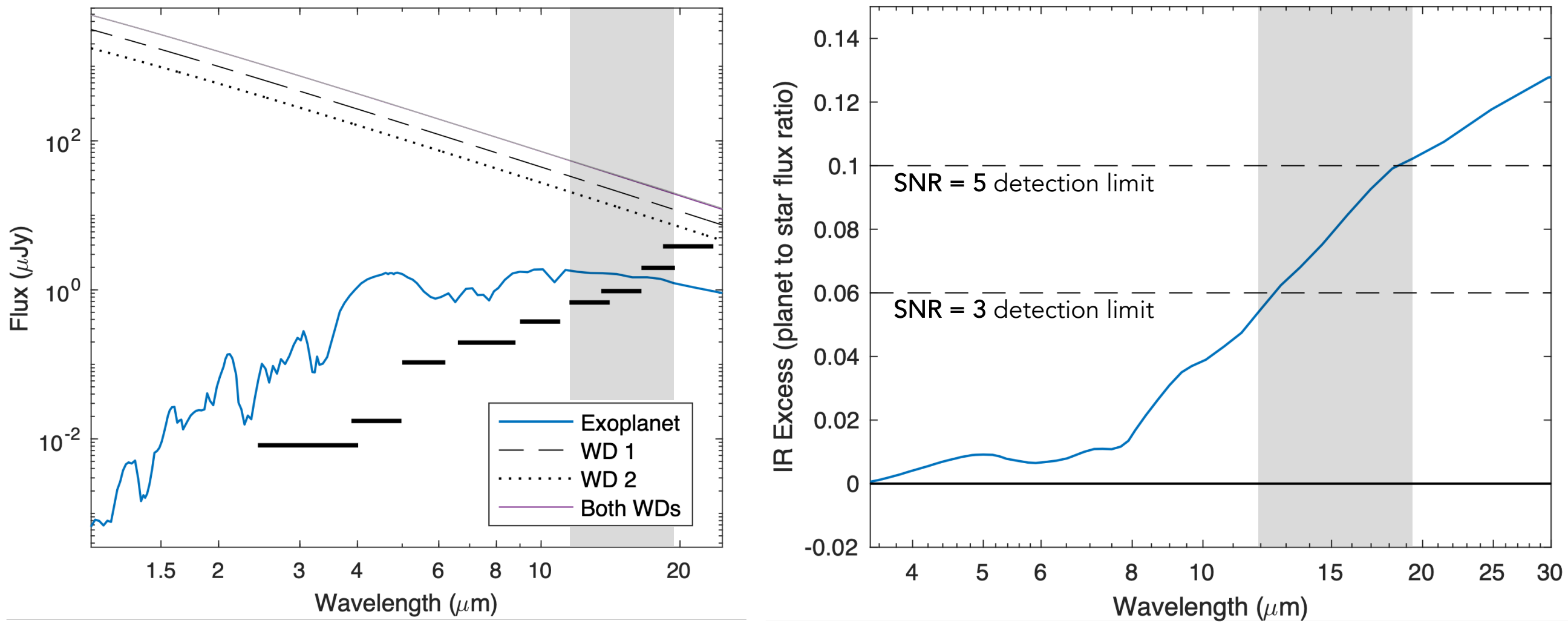

**Figure 2.** Simulations show IR excess detection of giant planets in DWD systems is achievable with JWST MIRI for a ZTFJ2243-like system. Left: SED of a 7 $M_{\mathrm{Jup}}$ exoplanet (blue line), and the individual flux contributions of the WDs in the ZTFJ2243 system (dotted and dashed lines), placed hypothetically at a distance of 150 pc. The combined and blended flux of these two WDs is shown in the top solid line. The black horizontal bars are the sensitivity limit (SNR = 3, 20 minutes) of MIRI in various imaging bands; sources with emission above the black bars are detectable if viewed in isolation. Right: in addition to sufficient MIRI imaging sensitivity, as illustrated on the left, sources not imaged in isolation (but in this case blended with the DWD spectrum) must produce at least 6% excess to be detectable (SNR = 3). For this test case, the planet produces an IR excess of approximately 6% in the system's combined SED—sufficient for an SNR $\approx$ 3 detection using existing JWST/MIRI broadband photometry methods. Gray-shaded regions in both panels indicate the optimal MIRI bands for exoplanet detection in this test case (F1280W, F1500W, and F1800W).

We also note that detecting exoplanets around DWDs requires accurate models for both WDs in the system; in particular, the temperatures and radii (or equivalently, a surface gravity and mass) of both stars must be constrained prior to the MIRI IR excess measurements. While some DWD systems are already sufficiently well characterized, many are not and would require additional observations to constrain their stellar parameters before meaningful searches for IR excesses can be performed.

A general characteristic of short-orbit DWDs is that they tend to be significantly lower mass than solitary WDs that have not undergone mass transfer. Lower-mass WDs have larger radii than their higher-mass counterparts, and thus larger emitting areas. For a given exoplanet, this means that the fractional IR excess it produces is smaller around these lower-mass (larger-radius) WDs than it would be around higher-mass (smaller-radius) ones. Combined with the fact that there are two WDs contributing to the continuum, this implies that the IR excess technique is generally limited to detecting more massive planets in DWD systems than would be possible around solitary WDs. Nevertheless, as we have demonstrated, detection of exoplanets around DWDs with the IR excess technique is still feasible for massive exoplanets.

## 3. Results

In the previous section, we established the methods for determining GW RV and IR excess sensitivity to DWD exoplanets. Figures 1 and 2 show overlapping parameter space accessible to both GW RV detections with LISA and IR excess detections with JWST, allowing a subset of planets to be independently confirmed or refuted. However, a substantial region of parameter space remains where planets lie beyond the IR sensitivity range will require alternative follow-up strategies. N. Tamanini & C. Danielski (2019) showed that astrometry probes a similar region of parameter space to GW DWD exoplanet detection. In this section, we build on this by incorporating astrometry to characterize the joint GW+EM parameter space and identify the most effective EM methods for planet confirmation. Our methods used for estimating astrometric sensitivity are described in the Appendix.

In Figure 3, we compare the regions of parameter space, planet mass versus the orbital period, observable at the SNR $> 5$ level for LISA GW RV, JWST IR excess, and HWO/Gaia astrometry. We emphasize again that the residual SNR is being used as a proxy for the companion SNR in the LISA curves shown. Because the JWST IR excess method only depends on the inherent thermal emission from the planet and not the planet's separation or orbital period, the overlap region in time with LISA's sensitivity is due entirely to the range in orbital periods where LISA matches JWST's performance, $\simeq 3 \times 10^{-3}$ yr $\approx$ 1 day for this source, and LISA's lifetime. In order to provide more context to this comparison, we include estimates of the astrometric performance of HWO and Gaia. We refer to J. Amiaux et al. (2025) for HWO's sensitivity, and use Gaia's predicted performance for its extended mission lifetime or DR5.[9]

As can be seen from Figure 3, the projected astrometric sensitivity of future facilities such as the HWO (National Academies of Sciences, Engineering, and Medicine 2021) is well matched to much of the same planetary parameter space probed by LISA. However, astrometric monitoring campaigns can be time-intensive and require long temporal baselines, whereas IR excess confirmation generally requires only a single epoch of observation. When feasible, IR excess follow-up is therefore significantly more observationally efficient.

In addition to IR excess and astrometric confirmation being essential for validating LISA-discovered exoplanet candidates, detecting a given planet with as many independent methods as possible, whether EM or GW, provides the strongest constraints on its orbital architecture and physical properties. Such multimessenger characterization enables a more complete and self-consistent understanding of these highly unusual systems. At present, only a small number of circumbinary planets are known, and none have been confirmed around DWDs (although PSR B1620-26 (AB)b orbits a pulsar/WD binary S. E. Thorsett et al. 1993). Constraining the occurrence rate and properties of these systems will substantially advance our understanding of post-main-sequence planetary evolution (D. Veras 2021), particularly in close binary environments.

[9] https://www.cosmos.esa.int/web/gaia/science-performance#astrometric%20performance

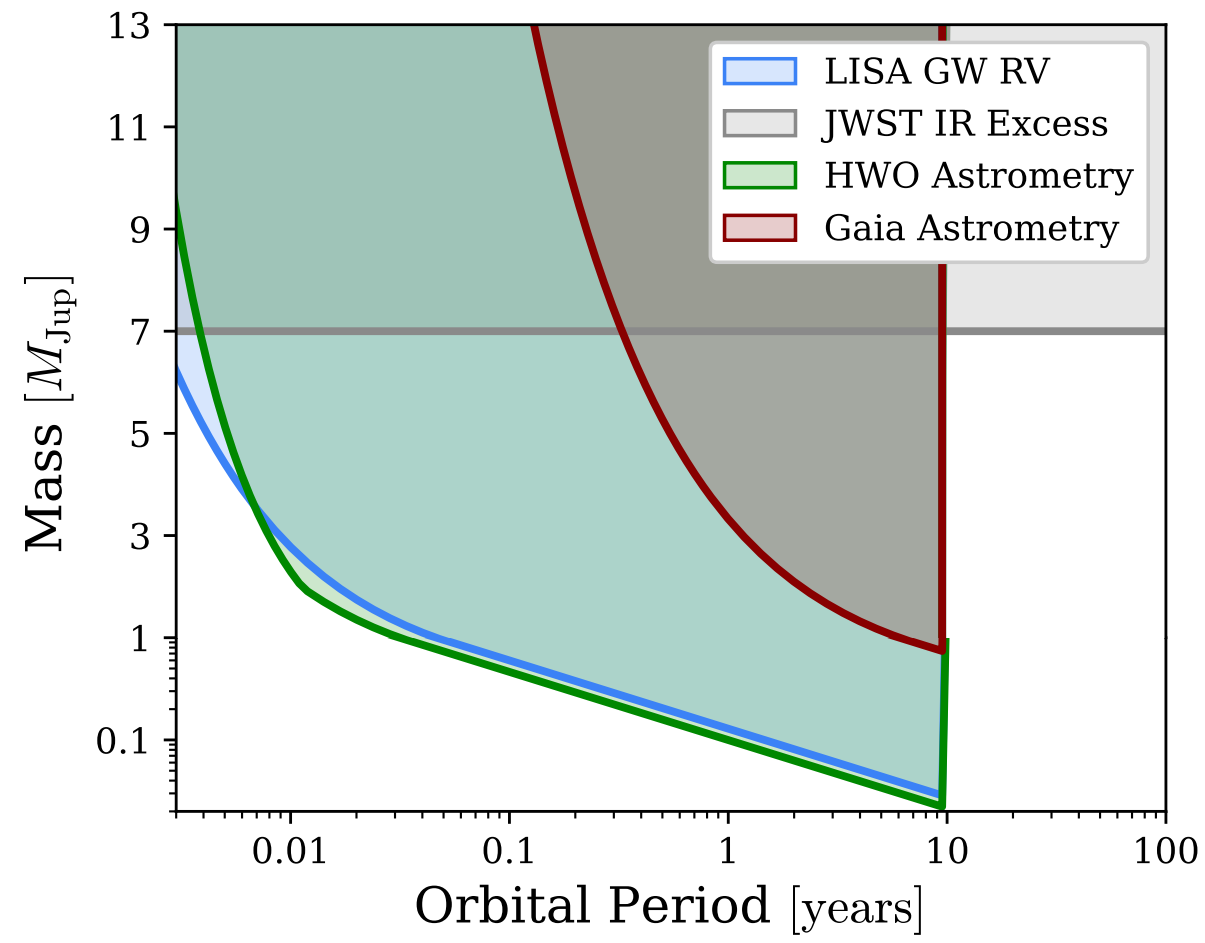

**Figure 3.** Joint exoplanet discovery space using multimessenger methods and different observatories for a ZTFJ2243-like system at a distance of 150 pc. Equation (5) is used to set the boundary of the LISA region, while Equation (A2) is used to set the HWO region for detecting exoplanets with astrometric shifts. An SNR = 5 is used for all curves, and we truncate the LISA and astrometric measurements to LISA's extended lifetime (10 yr). If HWO lasts longer than the 10 yr of LISA's extended lifetime, then its sensitivity would reach smaller planet masses and longer orbital periods than shown here. Gaia operated for approximately 10 yr, and its data is still being processed. The period turnover in the HWO curve is a consequence of Equation (A4). Note the transition on the $y$-axis from logarithmic to linear scale, which emphasizes the LISA and JWST overlap at higher mass values, and the power-law scaling of LISA and HWO at lower mass values.

We also investigated how sensitivity scales with distance. In Figure 4, we compare the astrometric reach of HWO and Gaia with the sensitivity of LISA to systems similar to ZTFJ2243. We find that LISA becomes more sensitive than HWO at a distance greater than 197 pc and may detect planets with masses greater than 0.17 $M_{\rm Jup}$ at that distance. We exclude IR excess detection from this plot, as shorter wavelengths (e.g., JWST NIRCam rather than MIRI) become necessary at larger distances, and the scaling is complex and nonlinear, requiring detailed modeling beyond the scope of this work.

We also find that LISA may detect planets with mass $M_p > 7\,M_{\rm Jup}$ at the Galactic center (8230 pc) about ZTFJ2243-like DWDs. We note that the astrometric signal was calculated ignoring Galactic extinction, so the HWO and Gaia estimates to distances out to the Galactic center are optimistic. However, because GWs pass losslessly through diffuse gas distributions, another major benefit of GWs over EM waves, the LISA signal suffers no additional diminution.

## 4. Discussion

Figure 1 shows that, when comparing models with and without an exoplanet, the residual SNR can exceed typical detection thresholds for increasing exoplanet mass and orbital period. However, this analysis does not fully account for parameter uncertainties and covariances. In GW literature, "effectual" models reproduce the observed signal well, whereas "faithful" models both match the signal and accurately represent its underlying physics (T. Damour et al. 1998). This study does not assess whether a DWD-only model can effectually reproduce a GW signal that includes an exoplanet. For example, uncertainties in DWD-only model parameters may mask the exoplanet signal, making detection more challenging than presented here.

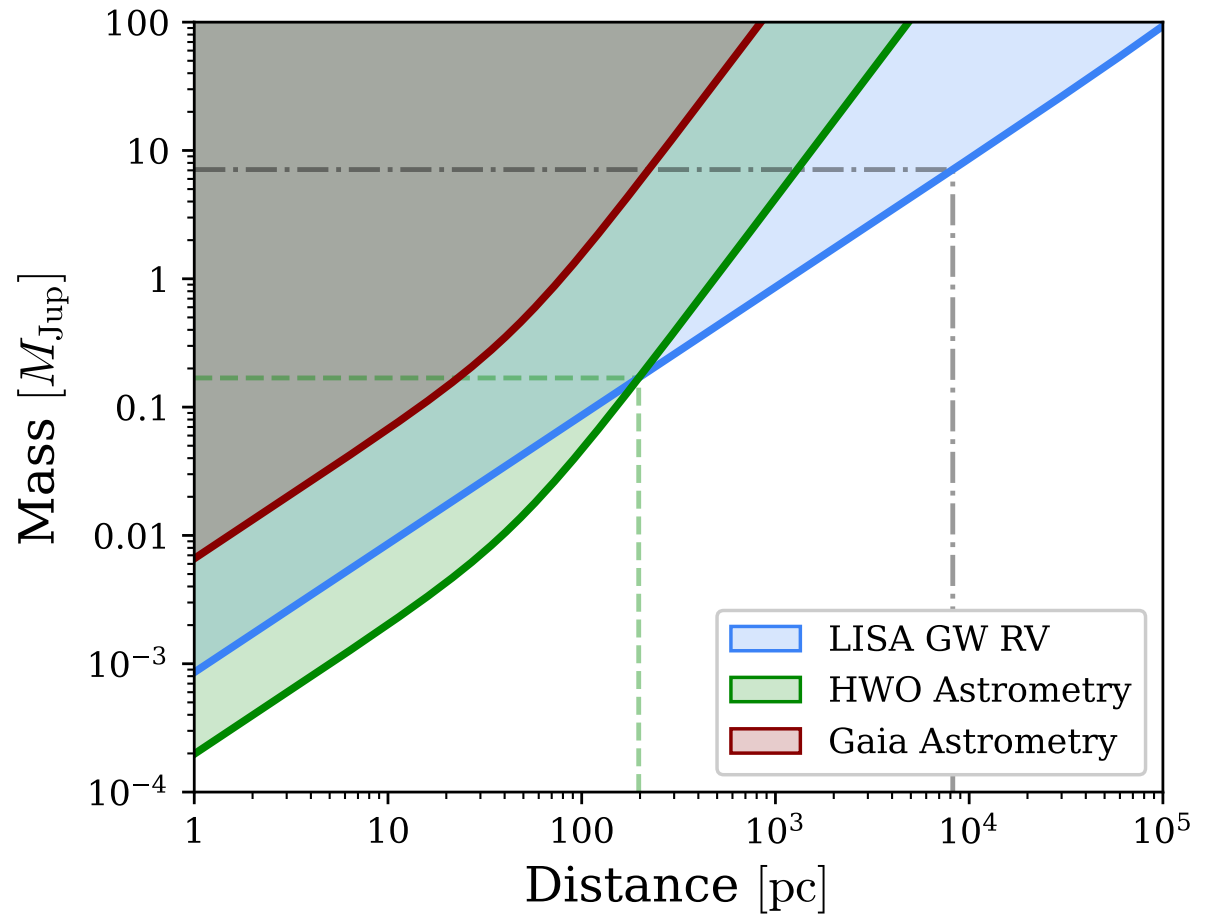

**Figure 4.** Planet mass vs. system distance discovery space, observable by LISA, HWO, and Gaia for a ZTFJ2243-like system assuming a 1 yr planet orbit. Green dashed lines indicate the distance beyond which LISA is more sensitive than HWO; we note that LISA is always more sensitive than Gaia in this case. The dotted–dashed curve indicates the distance to the Galactic center, 8230 pc. An SNR = 5 is used for all curves. The $D^{-1}$ to $D^{-2}$ turnover in the astrometric methods occurs when the astrometric error of the target source begins to dominate in Equation (A5).

Similarly, we do not yet understand whether faithful DWD +exoplanet models can also be effectual in modeling DWD-only systems that include other physical processes (e.g., mass transfer between the WDs). These effects may increase the likelihood of false positives in exoplanet detection, making EM detection of GW candidates critical.

These considerations highlight two key priorities for future work that will aid the success of the LISA mission. First, more robust studies of data analysis and parameter recovery using models that include effects such as exoplanets are needed to assess their effectualness and faithfulness. For example, future work could compute the mismatch between different DWD-only and DWD+exoplanet models (T. Damour et al. 1998; Y. Pan et al. 2011), or perform full Bayesian parameter estimation searches (e.g., M. L. Katz et al. 2022). This will help determine how close to the limits presented in this work can be realistically achieved with LISA.

Second, although we have emphasized LISA GW observations as an effective survey for identifying DWD exoplanet candidates, which can then be confirmed via IR excess or astrometric detection, these EM methods may also identify such systems prior to LISA. For example, Gaia astrometry may still identify candidate systems that are later confirmed to have DWD hosts. Additionally, the IR excess method explored here could be applied now using JWST to conduct targeted surveys for companions around known DWDs, particularly if the planets are young. This could motivate targeted near- and mid-IR surveys of known DWDs.

At present, the sample of EM-identified DWDs remains small, and many systems lie at relatively large distances from Earth, limiting (but not excluding) sensitivity to planetary companions. As a result, it is unclear whether such surveys would yield detections. This uncertainty is compounded by the poorly constrained occurrence rates of planetary systems in these environments, particularly for post-main-sequence systems where the host stars are stellar remnants. Nevertheless, any detection of a planetary companion to a DWD via EM methods would be highly valuable. Even if the host DWD is not in an orbit tight enough to generate GWs in the LISA sensitivity range, discovering such a circumbinary would help inform population synthesis models, occurrence rates, and the stellar-planetary evolution channels that yield such systems.

However, if such a pre-detected system does host a tight-orbit DWD generating GWs in the LISA band, then this discovery would be especially valuable as a benchmark for targeted GW searches, in the same way the currently known LISA DWD systems are E. Finch et al. (2023). Having known targets will better inform the development and characterization of both the GW models and the complex LISA search algorithms that could then validate these systems. Ultimately, this would then improve the chances of discovering the unknown circumbinaries.

## 5. Conclusion

We have studied the potential exoplanet discovery space for circumbinary DWD systems, using multimessenger observations with LISA GW, RV, JWST IR excess, and HWO/Gaia astrometry.

Using a ZTFJ2243-like DWD system placed at a fiducial distance of 150 pc as a test case, we find the following:

1. LISA GW RV observations can detect a single Jupiter-mass exoplanet across orbital periods of $\sim$0.05–10 yr with an SNR $>$ 5. Sub-Jovians may be detectable on $\sim$1–10 yr orbits in this ZTFJ2243-like DWD system.
2. We present a novel method for detecting IR excess from exoplanets around DWDs, extending its previous application to solitary WD systems.
3. For this method and the fiducial ZTFJ2243-like system, we find a minimum detectable exoplanet mass of 5 $M_{\rm Jup}$ (SNR = 3).

We explore the overlap in parameter space between GW and EM detection of DWD exoplanets:

1. We find that IR excess and LISA GW RV detections overlap in sensitivity to DWD exoplanets, but with GW probing lower-mass planets and IR excess accessing a broader range of orbital periods.
2. We demonstrate that Gaia or HWO astrometry can detect most GW-identified exoplanets that are inaccessible to the IR excess method. Overall, these methods have comparable sensitivity, with LISA outperforming Gaia, and HWO astrometry exceeding LISA's sensitivity limits.
3. At distances $\gtrsim$200 pc LISA outperforms all methods in its potential reach, since GW amplitude is inversely proportional to the source distance. At shorter distances, the astrometric method has the same distance dependence, but then turns over to being inversely proportional to the squared source distance as the astrometric error of the DWD target begins to dominate.
4. We find that, while EM techniques are unlikely to efficiently survey large populations of DWDs to identify such exoplanets, GW candidate detection followed by targeted EM confirmation provides a feasible pathway to confirming a population of DWD exoplanets, should such a population exist. However, having pre-discovered systems from JWST or Gaia would greatly inform the development of the models and methods for detecting these systems with LISA.

And finally, to summarize the power-law relationships:

1. Equation (5) is the power law explicitly demonstrated in Figure 1, for fixed $\mathcal{M}$, $D$, $T_{\rm obs}$, $f_{\rm GW,0}$, and $i$.
2. Equation (6) is the more general power law made under the small phase modulation index assumption, which is expected for the majority of the LISA-regime systems.
3. Equation (A5) gives the HWO power law modified for a DWD circumbinary. Gaia's performance is scaled to this.

As we enter the era in which LISA can identify such planetary candidates through GW signals, IR excess measurements with JWST provide a viable and compelling avenue for EM follow-up and potential confirmation. Joint GW/EM detections, particularly for the first few planetary candidates identified by LISA, will be critical for establishing that the observed GW signals correspond to genuine planets rather than false positives. Moreover, if some candidates do prove to be false positives, these joint detections will enable us to quantify the false-positive rate.

A multimessenger exoplanet would be a historic discovery, and a DWD circumbinary would help to provide key insights into planetary and stellar evolution and into a part of the exoplanet census not yet probed. JWST, LISA, and HWO represent both current and future major space-based astrophysics missions that may enable such a discovery.

## Acknowledgments

The authors would like to give special thanks to Ravi Kopparapu for helping to form this group and his involvement throughout the course of this work. Additionally, the authors would like to thank Joe Renaud, Tyson Littenberg, J. Ira Thorpe, and John Baker for helpful conversations that helped guide this study, and the anonymous reviewer for their feedback.

The material presented here is based upon work supported by NASA under award No. 80GSFC24M0006. The authors also acknowledge support from the NASA Goddard Space Flight Center "Sellers Exoplanet Environments Collaboration" (SEEC), which is funded by the NASA Planetary Science Division's Internal Scientist Funding Model.

## Materials Availability

Code and materials to reproduce the calculations and the plots in this paper are publicly available under an open-source Apache-2.0 License: doi: 10.5281/zenodo.22129575.

## Author Contributions

1. C.M. was the principal investigator for this project, acquired funding, performed the LISA analysis and investigation, and helped with the writing.
2. M.A.L. performed the JWST analysis and investigation, and helped with the writing and editing.

3. S.C.N. performed the analysis comparing HWO and Gaia astrometry to the LISA and JWST results, and assisted with some of the writing.
4. V.K. assisted in conceptualization, funding acquisition, and writing—review and editing.

*Software.* `NumPy` (C. R. Harris et al. 2020), `SciPy` (P. Virtanen et al. 2020), `Matplotlib` (The Matplotlib Development Team 2025), `LISAOrbits` (J.-B. Bayle et al. 2024b), `LISAGWResponse` (J.-B. Bayle et al. 2024a), `PyTDI` (M. Staab et al. 2023).

## Appendix
## HWO Astrometry

The HWO is the next flagship mission being developed in NASA's Astrophysics portfolio and whose primary science mission is to discover and observe relatively nearby exoplanets in the habitable zone of their stellar system. Current single astrometric requirements for HWO are set so that a $M_p = 1\,M_\oplus$ exoplanet located at separation $a_p = 1$ au around a $M_\star = 1\,M_\odot$ star at distance $D = 10$ pc would have an SNR of at least 6 over $N_{\rm obs} = 100 \lesssim 1$ hr pointings with a single measurement uncertainty of $\sigma_{1\,\rm meas} = 0.5\,\mu$as (J. Amiaux et al. 2025):

$$\mathrm{SNR}_{\sim\odot} = 6\left(\frac{M_p}{M_\oplus}\right)\left(\frac{a_p}{1\ \mathrm{au}}\right)\left(\frac{M_\star}{M_\odot}\right)^{-1} \times \left(\frac{D}{10\ \mathrm{pc}}\right)^{-1}\left(\frac{N_{\rm obs}}{100}\right)^{1/2}\left(\frac{\sigma_{1\,\rm meas}}{0.5\ \mu\mathrm{as}}\right)^{-1}. \tag{A1}$$

The HWO SNR relation assumes a flux from a Sun-like star ($\mathrm{SNR}_{\sim\odot}$) with radius $R_{\rm HWO} = 0.997 R_\odot$, temperature of $T_{\rm HWO} = 5660$ K, at a distance $D_{\rm HWO} = 10$ pc. Because our WDs are significantly hotter, $T_A = 26{,}300$ K and $T_B = 19{,}200$ K, are significantly smaller, $R_A = 0.0298 R_\odot$ and $R_B = 0.0275 R_\odot$, and significantly farther, $D = 150$ pc, we must adjust this figure for our ZTFJ2243-like system ($\mathrm{SNR}_{\rm DWD}$; K. B. Burdge et al. 2020).

The measurement uncertainty, $\sigma_{1\,\rm meas}$, is the quadrature sum of different errors or uncertainties (see Figure 1 of J. Amiaux et al. 2025). The only one that concerns us here is the astronomic error of the target source, $\sigma_{\rm DWD}$, here that of the DWD system. This uncertainty is proportional to $N_\gamma^{-1/2}$, where $N_\gamma$ is the number of photons incident to the detector (J. Amiaux et al. 2025). Since $N_\gamma \propto F_\lambda$, where $F_\lambda$ is the flux per unit wavelength of the incident light and $F_\lambda \propto R_\star^2 B_\lambda(T)/D^2$, where $R_\star$ is the radius of the star, $B_\lambda(T)$ is the spectrum of a blackbody at temperature $T$, and $D$ is the distance to the source, then we must multiply Equation (A1) by a term that compensates for the differences in color and size:

$$\sigma_{\rm DWD} = \sigma_{\sim\odot}\left[\frac{R_{\rm HWO}^2 B_\lambda(T_{\rm HWO})/D_{\rm HWO}^2}{(R_A^2 B_\lambda(T_A) + R_B^2 B_\lambda(T_B))/D^2}\right]^{1/2} = \sigma_{\sim\odot}\left[\frac{R_{\rm HWO}^2 B_\lambda(T_{\rm HWO})}{R_A^2 B_\lambda(T_A) + R_B^2 B_\lambda(T_B)}\right]^{1/2}\left(\frac{D}{D_{\rm HWO}}\right), \tag{A2}$$

where $\sigma_{\sim\odot} \simeq 0.03\,\mu$as is the astrometric error for the reference HWO system. We note that this dependence only matters if $\sigma_{\rm DWD}$ dominates the total error budget, $\sigma_{1\,\rm meas}$. We leave the other error contributions unchanged from J. Amiaux et al. (2025) and so

$$\sigma_{1\,\rm meas} = \sqrt{\sigma_{\rm DWD}^2 + \cdots}, \tag{A3}$$

where the ellipsis implies the rest of the quadratic sum.

The baseline figure of merit includes 100 1 hr long observations, amounting to approximately 1% of a year. If the planet's period is shorter than 100 hr, then fewer observations will be possible, diminishing the astrometric accuracy of the measurement. In summary,

$$N_{\rm obs} = \begin{cases} 100, & 100\ \mathrm{hr} < P, \\ \frac{P}{1\ \mathrm{hr}}, & 6\ \mathrm{hr} < P < 100\ \mathrm{hr}. \end{cases} \tag{A4}$$

We note that in order to adequately resolve a full cycle of an orbit, a minimum of $\simeq 6$ observations would be needed, meaning that HWO as we model it here would be limited to $P > 6$ hr. When the planet's period is between 6 and 100 hr, this condition has the consequence of introducing a factor of $P^{1/2}$ in the expression for $\mathrm{SNR}_{\rm DWD}$ and results in an overall dependence of $P^{-7/6}$ in Equation (A5).

Putting this in terms of planet mass versus orbital period,

$$M_p = 1\,M_\oplus\left(\frac{P}{1\ \mathrm{yr}}\right)^{-2/3}\left(\frac{M_{\rm DWD}}{M_\odot}\right)^{2/3}\left(\frac{D}{10\ \mathrm{pc}}\right) \times \left(\frac{\mathrm{SNR}_{\rm DWD}}{6}\right)\left(\frac{N_{\rm obs}}{100}\right)^{-1/2}\left(\frac{\sigma_{1\,\rm meas}}{0.5\ \mu\mathrm{as}}\right). \tag{A5}$$

We note that $\sigma_{1\,\rm meas}$ is now a function of the DWD system's properties via Equations (A2) and (A3).

Gaia DR4 is expected to have approximately 300 times poorer precision (J. Amiaux et al. 2025), so we multiply HWO's $\mathrm{SNR}_{\rm DWD}$ by 1/300 to approximate Gaia's performance.

## ORCID iDs

Casey McGrath https://orcid.org/0000-0002-6155-3501
Mary Anne Limbach https://orcid.org/0000-0002-9521-9798
Scott C. Noble https://orcid.org/0000-0003-3547-8306
Vincent Kofman https://orcid.org/0000-0002-5060-1993